\documentclass{mhd}     
\usepackage{graphicx}
\usepackage{booktabs}
\usepackage{rotating}
\usepackage[font=small,labelfont=bf]{caption}
\usepackage{hyperref}
\usepackage[version=4]{mhchem}  
\usepackage{amssymb, amsmath, mathtools, commath}
\usepackage{color}
\usepackage{yfonts}
\usepackage{siunitx}
\usepackage{float}
\usepackage{subfigure}
\usepackage{comment}  

\title{Effects of Current Distribution on Mass Transport in the Positive Electrode
of a
Liquid Metal
        Battery}	
\author{P.~Personnettaz\inst{*}, S.~Landgraf, M.~Nimtz,
        N.~Weber, T.~Weier}     
\institute{Helmholtz--Zentrum Dresden -- Rossendorf, Bautzner Landstr. 400,
        Dresden, Germany}
\dedication{*Corresponding author: p.personnettaz@hzdr.de}

\begin{document}

\maketitle
\begin{abstract}%
Liquid metal electrodes are one of the key components of
different electrical energy storage technologies. The
understanding of transport phenomena in liquid electrodes is mandatory in
order to ensure efficient operation.
In the present study we focus our attention on the positive electrode of a
Li$||$Bi liquid metal battery.
Starting from a real experimental setup, we numerically investigate charge
transfer in the molten salt electrolyte and mass transport in the
positive electrode.
The two phenomena are tightly coupled, because the current
distribution influences the concentration field in the
positive electrode. The cell is studied during charge, when compositional
convection becomes apparent. First results of
compositional convection from an inhomogeneous current distribution are
presented, highlighting its capability to affect the flow in the positive
electrode and the cell performance. \\
\end{abstract}

\section*{Introduction.}
\label{sec:intro}
Electrical energy storage (EES) is recognized as to be essential for a broader
deployment of intermittent renewable sources in the electricity production.
Liquid metal batteries (LMBs) \cite{kim2012liquid, kelley2018fluid},
sodium-sulfur \cite{hueso2013high} and liquid
metal displacement batteries (LDBs) \cite{yin2018faradaically} are promising
EES candidates. They all have in common a key component, the liquid metal
electrode (LME). The fully fluid nature of LMEs guaranties simple construction
(scale
up) of the cell and extended lifetime \cite{li2016liquid}. Simultaneous charge,
heat,
mass and momentum transfer takes place in these electrodes and  affects their
electrochemical behavior.  The actual knowledge of these phenomena is far from
being comprehensive, despite the relatively simple geometry and chemistry.
The high operating temperatures and the chemical aggressive environment limit
the experimental study to few measurable quantities. Numerical studies based
on continuum mechanics allow to infer the physics and to understand the
relevant transport mechanisms. 

The presence of high current density across conducting fluids (liquid metals) has
promoted the investigation of multiple magnetohydrodynamic (MHD) phenomena \cite{kelley2018fluid}:
liquid metal columns crossed by electrical current may be subject to the Tayler instability 
\cite{weber2014current, weber2015influence}; interface instabilities can be excited by Lorentz forces \cite{weber2017sloshing};
diverging current distributions may lead to electrovortex flows \cite{weber2015influence,ashour2018competing,weber2018electromagnetically,weber2020numerical, herreman2019numerical}. 
Looking at non-uniform current distributions from a different perspective, motivated by an ongoing experiment and a previous work
\cite{personnettaz2019mass}, in this study we want to investigate the effect
on mass transport
of an inhomogeneous current profile on the charging phase of a \ce{Li||Bi}
liquid metal battery.

In the experiments published so far
the negative electrode has different shapes (spherical droplet
\cite{weier2017liquid}, cylindrical porous matrix \cite{wang2014lithium} or
spiral \cite{personnettaz2019mass}). In all of these setups the negative active surface
is smaller with respect to the positive electrode-electrolyte interface. This
affects the primary current distribution in the electrolyte and the mass
flux at the top interface of the positive electrode.
During charge, \ce{Li} is removed (negative mass flux) from the
top interface of the positive electrode (electrorefining) through a process of
\ce{Li} reduction and migration in the electrolyte layer. This phenomenon
generates denser fluid at the top part of the positive electrode.
The density stratification is unstable and the fluid parcels sink down,
thereby producing a compositional convective flow \cite{kelley2018fluid}.
In the previous work we have assumed a uniform mass flux
\cite{personnettaz2019mass}, here we study if
this approximation is justified, and what effect it has on the estimation of the
cell performance.

\section{Experimental setup.}
\begin{figure*}[t!]
        \centering
        \begin{subfigure}[]
                \centering
                \includegraphics[width=0.48\linewidth]{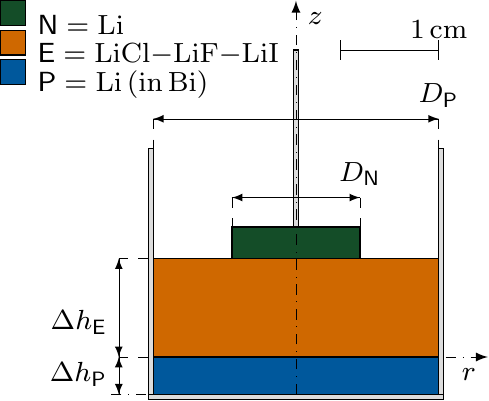}
        \end{subfigure}%
        \begin{subfigure}[]
        \centering
        \includegraphics[width=0.48\linewidth]{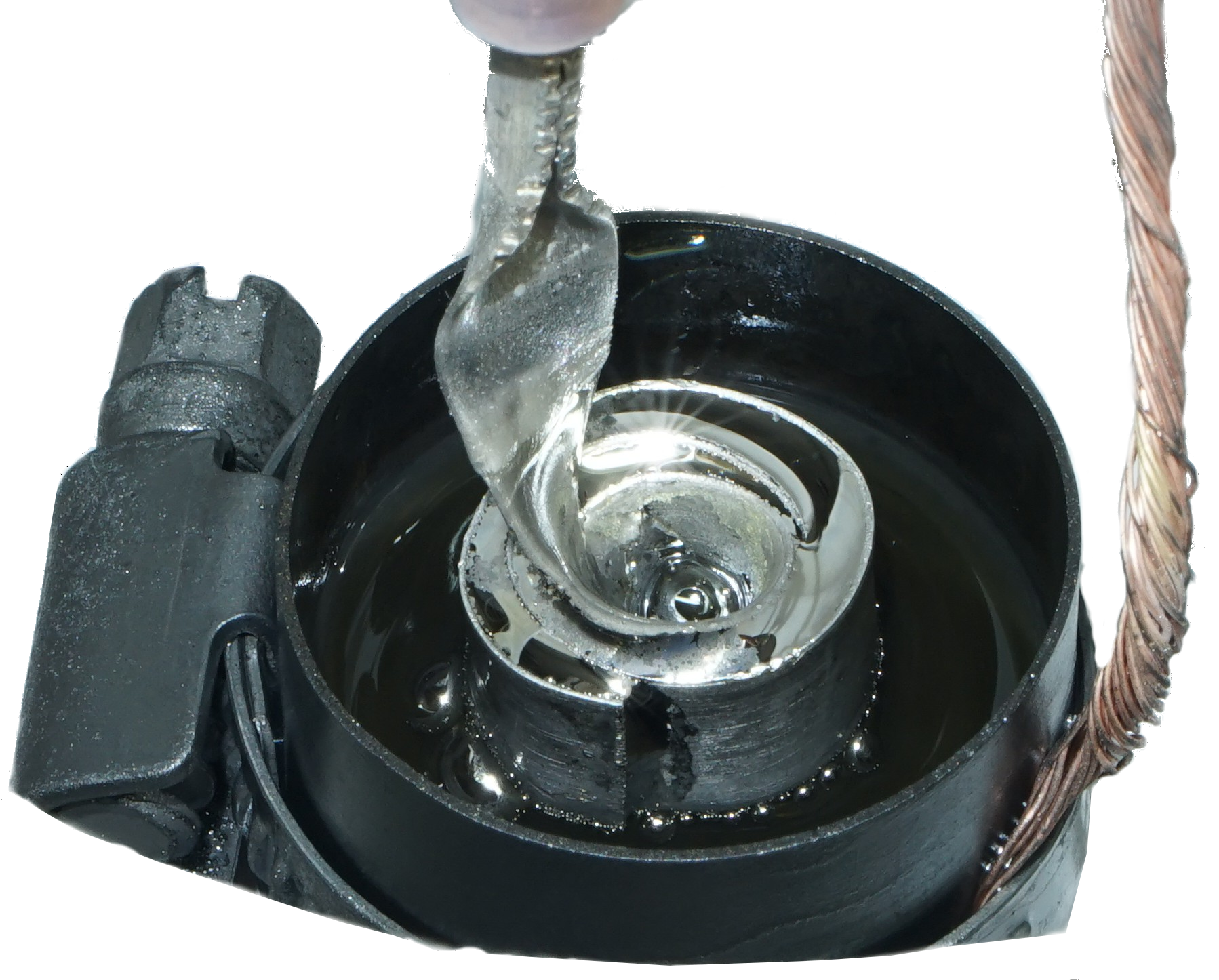}
        \end{subfigure}
        \caption{\label{fig:cell} (a) Geometry of the experimental
                \ce{Li||Bi} cell. (b)  Experimental \ce{Li||Bi} cell.}
\end{figure*}

The \ce{Li||Bi} cell is composed of three liquid layers stratified by
density, as shown in Fig.~\ref{fig:cell}a.
The positive electrode consists of a liquid alloy
(\ce{Li(in Bi)}) and it constitutes the lower layer. On top of that an
eutectic
\ce{LiCl-LiF-LiI} molten salt electrolyte floats. Both layers are contained in
a tantalum vessel, in order to minimize side reactions. The top lithium
(\ce{Li}) negative electrode is laterally confined in a nickel spiral, as
shown in Fig.~\ref{fig:cell}b.
The concentration cell is open and immersed in a controlled
\ce{Ar} atmosphere, in order to ensure safety and optical access
\cite{weier2017liquid}. A heating plate placed on the bottom of the vessel
keeps the cell core temperature of about \SI{460}{\celsius}.
The geometrical parameters and the relevant material properties are compiled
in
Tab~\ref{tab:operatingParamters}.
\begin{table}[h!]
        \caption{Relevant geometrical parameters of the \ce{Li||Bi}
        LMB; thermodynamic and transport properties of the \ce{Li(in Bi)} alloy
        and	\ce{LiCl-LiF-LiI} molten salt electrolyte at $T = $~\SI{460}{\celsius}
        \cite{fazio2015handbook, ohse1985handbook, temnogorova1980determination,
        newhouse2014modeling, masset2007thermal}.}
        \label{tab:operatingParamters}
        \begin{center}
                \begin{tabular}{lcc}
                        \toprule
                        diameter of the inner vessel & $ D_\text{P}$ &
                        \SI{29}{\milli\meter}\\
                        diameter of the negative electrode & $ D_\text{N}$ &
                        \SI{13}{\milli\meter}\\
                        thickness of the molten salt layer & $ \Delta h_\text{E}$ &
                        \SI{10}{\milli\meter}\\
                        thickness of the positive electrode & $ \Delta h_\text{P}$ &
                        \SI{3.8}{\milli\meter}\\ \midrule
                        mixture density  & $\rho_\text{tot,ref}$ &
                        \SI{8.2e3}{\kilogram\per\cubic\meter}\\
                        compositional expansion coefficient  &
                        $\beta_\rho$ &
                        \SI{1.88e-3}{\cubic\meter\per\kilogram}\\
                        reference mass concentration of \ce{Li}  &
                        $\rho_\text{A,ref}$ &
                        \SI{9.045e1}{\kilogram\per\cubic\meter}\\
                        kinematic viscosity  & $\nu$ &
                        \SI{1.31e-7}{\square\meter\per\second}\\
                        diffusion coefficient of \ce{Li} in \ce{Bi} &
                        $\mathcal{D}_\text{AB}$ &
                        \SI{5e-9}{\square\meter\per\second}\\
                        molar mass of Lithium &
                        $\mathcal{M}_\text{A}$ &
                        \SI{6.941e-3}{\kilo\gram\per\mole}\\
                        \midrule
                        ionic conductivity of the electrolyte &
                        $\sigma_\text{el, E}$ &
                        \SI{2.71e2}{\siemens\per\meter}\\
                        \bottomrule
                \end{tabular}
        \end{center}
\end{table}

\section{Numerical model}
In order to study the effect of an inhomogeneous current density on the
\ce{Li}-concentration
distribution we first solve the electrodynamics
problem and then we apply the derived current profile as boundary
condition for
mass transport in the positive electrode. In order to simplify the model
and to reduce the numerical effort we use the axisymmetric approximation.
\subsection{Electrodynamics}
\begin{figure*}[t!]
        \centering
        \begin{subfigure}[]
                \centering
                \includegraphics[width=0.80\linewidth]{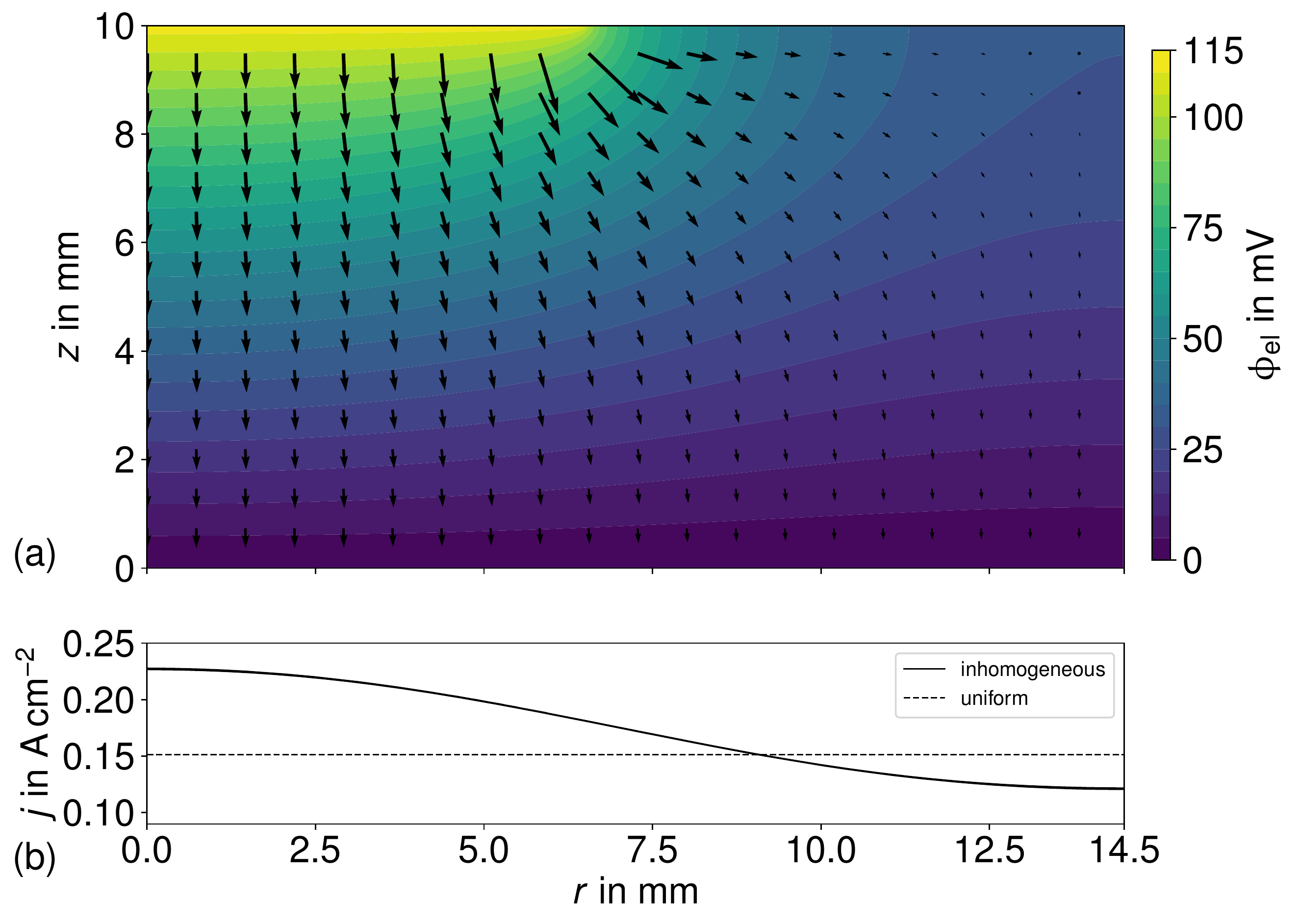}
        \end{subfigure}%

        \caption{\label{fig:phiEl} (a) Electric potential $\phi_\text{el}(r, z)$
        and current density $\textbf{j}(r, z)$
        distribution in the electrolyte. (b) Current density
                profile $j(r, z = 0)$ at the interface between electrolyte and positive
                electrode for $I =
                \SI{1}{\ampere}$ .}
\end{figure*}

The primary current distribution at the interface between the
electrolyte and the positive electrode $j(r, z = 0) = j (r)$ can be
numerically computed from the solution of the Laplace equation for the
electric potential $\phi_\text{el}$ in the electrolyte:
\begin{equation}
\nabla  \cdot \sigma_\text{el, E} \nabla \phi_\text{el} = 0 \ ,
\end{equation}
in which $\sigma_\text{el, E}$ is the ionic conductivity of the molten salt.
\begin{figure*}[t!]
        \centering
        \begin{subfigure}[]
                \centering
                \includegraphics[width=0.65\linewidth]{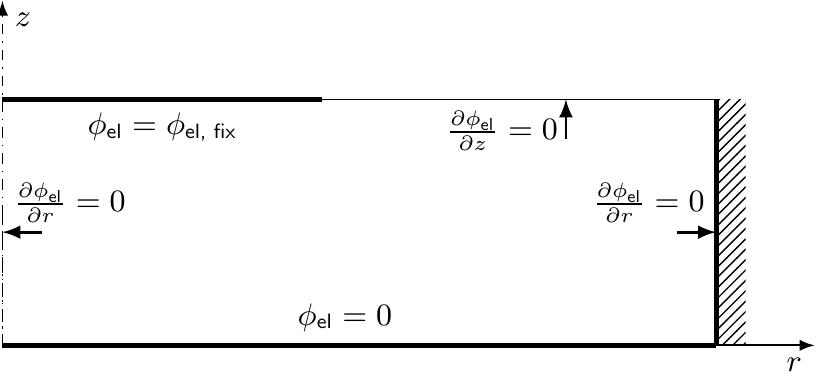}
        \end{subfigure}%
        \begin{subfigure}[]
                \centering
                \includegraphics[width=0.65\linewidth]{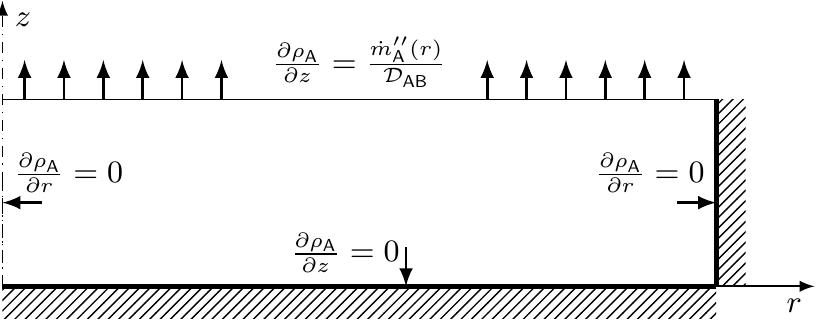}
        \end{subfigure}
        \caption{\label{fig:bc} (a) Electrolyte (axial symmetry),
        boundary conditions for the primary current distribution.  (b)
                Positive
                electrode (axial
                symmetry), boundary conditions for
                the mass transport equation.}
\end{figure*}
We assume a Dirichlet boundary condition for the interfaces between
electrolyte and liquid metals and a homogeneous Neumann boundary condition
elsewhere, as shown in Fig.~\ref{fig:bc}a. The current density $\textbf{j}$ is
derived from the Ohm law, $\textbf{j} = -\sigma_\text{el, E} \nabla
\phi_\text{el}$. The solution was computed with the Laplace solver of the free
and open source CFD library \texttt{OpenFOAM} and is presented in
Fig.~\ref{fig:phiEl}.
These results were compared with the primary current distribution in
the full cell and an excellent agreement was confirmed for the current density
at the electrolyte-positive electrode interface. Eq.~1 can
be solved analytically as well in the form of a infinite sum of Bessel
functions. We use a polynomial fit of the fifth order to interpolate the radial
dependence of the current profile $j(r) = \langle j\rangle \cdot
\sum^{5}_{i=0} a_{i}\cdot r^{i}$.
\subsection{Mass transport}
We study concentration and flow fields in the positive electrode with a single
phase-single region model.
The following assumptions are used: isothermal conditions; rigid and flat
interface with the molten salt electrolyte.
The concentration distribution is modeled with the advection-diffusion equation:
\begin{equation}
        \pd{\rho_\text{A}}{t} + \textbf{u}\cdot \nabla \rho_\text{A} = \nabla
        \cdot\mathcal{D}_\text{AB} \nabla\rho_\text{A}
\end{equation}
in which $\rho_{\text{A}}$ is the mass concentration of the component A
(\ce{Li}), $\textbf{u}$ is the
velocity field and
$\mathcal{D}_\text{AB}$ is the diffusion coefficient
\cite{bird2002transportphenomena}. The boundary conditions (BCs)
for mass transport are shown in Fig.~\ref{fig:bc}b. We consider two
configurations: a uniform (a) and an inhomogeneous (b) mass flux at the top
interface
$\dot{m}''_\text{A}$:
\begin{equation}
(a) \quad \dot{m}''_\text{A} =
\frac{\langle
j\rangle\mathcal{M}_\text{A}}{n_\text{el}\text{F}\mathcal{D}_\text{AB}}
\qquad (b) \quad
\dot{m}''_\text{A} (r) =
\frac{j(r)\mathcal{M}_\text{A}}{n_\text{el}\text{F}\mathcal{D}_\text{AB}} =
\frac{\langle
        j\rangle\mathcal{M}_\text{A}}{n_\text{el}\text{F}\mathcal{D}_\text{AB}}\sum^{5}_{i=0}
         a_{i}\cdot r^{i} \
\end{equation}
in which $\langle j\rangle$, $\text{F}$, $\mathcal{M}_\text{A}$, $n_\text{el}$,
are mean current density, Faraday constant, molar mass of \ce{Li} and  the
number of electrons transferred per ion ($n_\text{el} = 1$) respectively. The
mean current density is calculated with  the positive electrode area ($\langle
j
\rangle = \frac{4 \cdot I}{\pi D_\text{P}^2}$).
The velocity field $\textbf{u}$ is computed by solving the incompressible
Navier-Stokes equations in the Boussinesq approximation: the
continuity equation,
\begin{equation}
        \nabla\cdot\textbf{u} = 0 \ ,
\end{equation}
and the momentum equation,
\begin{equation}
        \label{Eq:Momentum2NavierStokes}
        \frac{\partial\mathbf{u}}{\partial t}
        +(\mathbf{u}\cdot\nabla)\mathbf{u}= - \frac{1}{\rho_\text{tot, ref}}\nabla p
        +\nu\nabla^2\mathbf{u} +
        \mathbf{g}(1-\beta_\rho(\rho_\text{A}
        -\rho_\text{A, ref})) \ ,
\end{equation}
in which $p$, $\rho_\text{tot, ref}$, $\nu$, $\mathbf{g}$, $\beta_\rho $ are
pressure, total mixture density, kinematic viscosity, gravitational
acceleration and
composition coefficient of volume expansion \cite{leal2007advanced}. The
density of the alloy is calculated with Vegard's law
\cite{vegard1921konstitution}. We apply a
no-slip condition at the solid boundaries and at the interface with the
electrolyte layer. We ensure the symmetry condition on the $z$-axis. The
numerical model is implemented in \texttt{OpenFOAM}, by extending
 the \texttt{buoyantBoussinesqPimpleFoam} solver.
The inhomogeneous current boundary condition generates an uneven current distribution in the liquid metal layer.
 The interaction of the current with its own induced magnetic field leads to a
 Lorentz force distribution with a non-vanishing rotational part. Such a force
 field can drive a fluid flow, known as electrovortex flow (EVF).
 In the work of Herreman et al. \cite{herreman2020solutal} it is shown that compositional convection exceeds EVF
 in a parameter range similar to the one studied. Furthermore, in our experiment the current profile at the interface is smoother with respect to the ones studied by Herreman et al.
 \cite{herreman2019numerical, herreman2020solutal}. Therefore, we expect an even weaker EVF.
 Due to these considerations, we neglect its contribution, and we assume that magnetohydrodynamic effects are negligible. In the case of an externally applied magnetic field, the validity of this assumption should be carefully verified \cite{weber2020numerical}.

\section{Results and discussion}
In this section we present first results on compositional convection in
the positive electrode under the effect of a uniform and an inhomogeneous mass
flux.
During charge the liquid alloy behaves like a viscous fluid cooled from the top
surface (e.g. evaporative cooling \cite{foster1965onset}). The mass flux
removes \ce{Li} from the \ce{Li (in Bi)} alloy, which generates continuously a
locally denser fluid. The phenomenon studied is intrinsically unsteady due to
the flux boundary condition. 

\begin{figure*}[t!]
        \centering
        \includegraphics[width=0.65\linewidth]{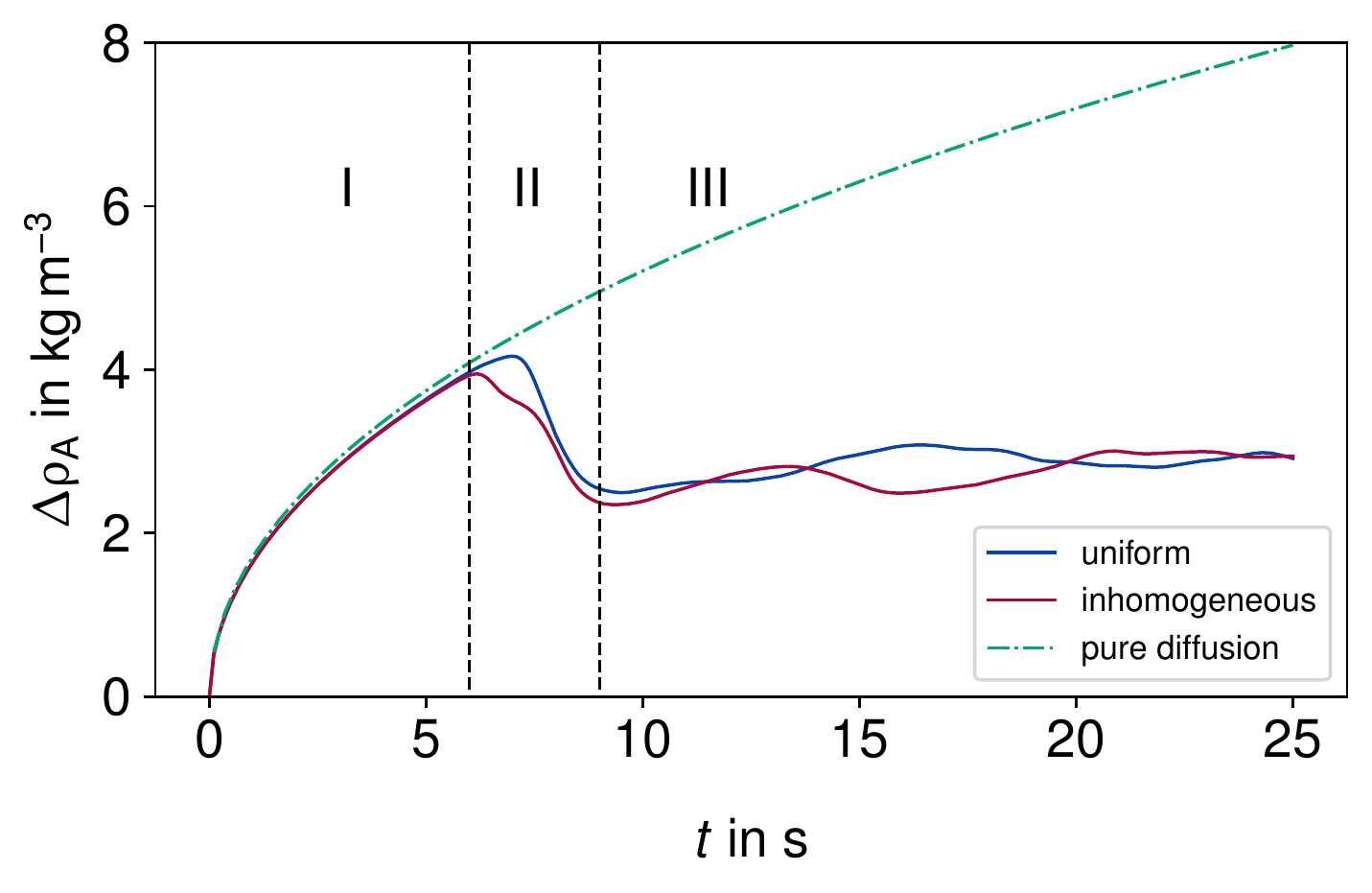}
        \caption{\label{fig:dRhoA} Evolution of the concentration difference
        between the bulk fluid and the interface for $I =
        \SI{1}{\ampere}$ ($\Delta \rho_{\text{A}}(t) =
         \langle \rho_{\text{A}} \rangle_{V}(t) - \langle
         \rho_{\text{A}} \rangle_{S}(t)$).}
\end{figure*}

\begin{figure*}[t!]
        \centering
        \begin{subfigure}[]
                \centering
                \includegraphics[width=1.2\linewidth]{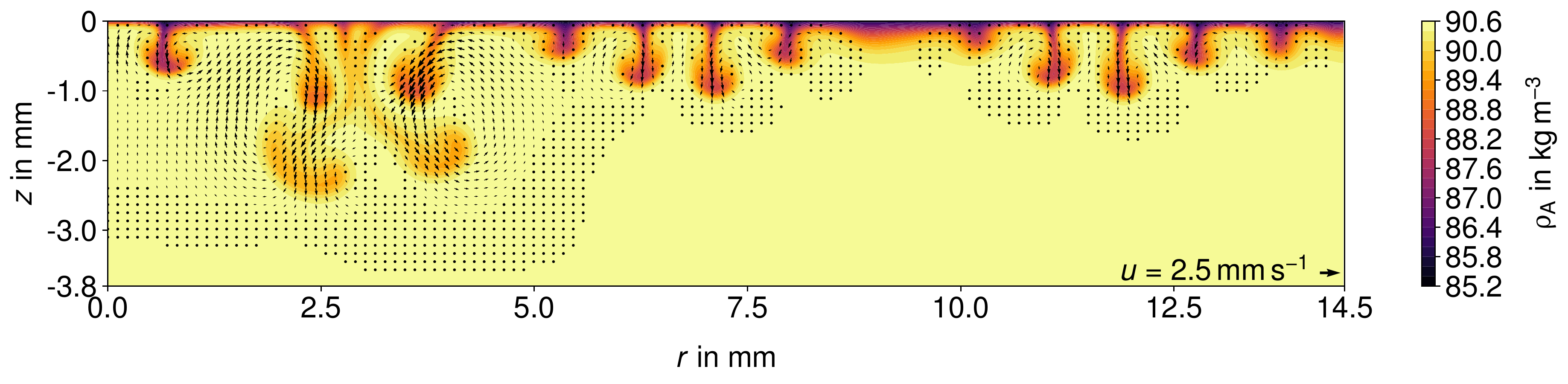}
        \end{subfigure}%
        \begin{subfigure}[]
        \centering
        \includegraphics[width=1.2\linewidth]{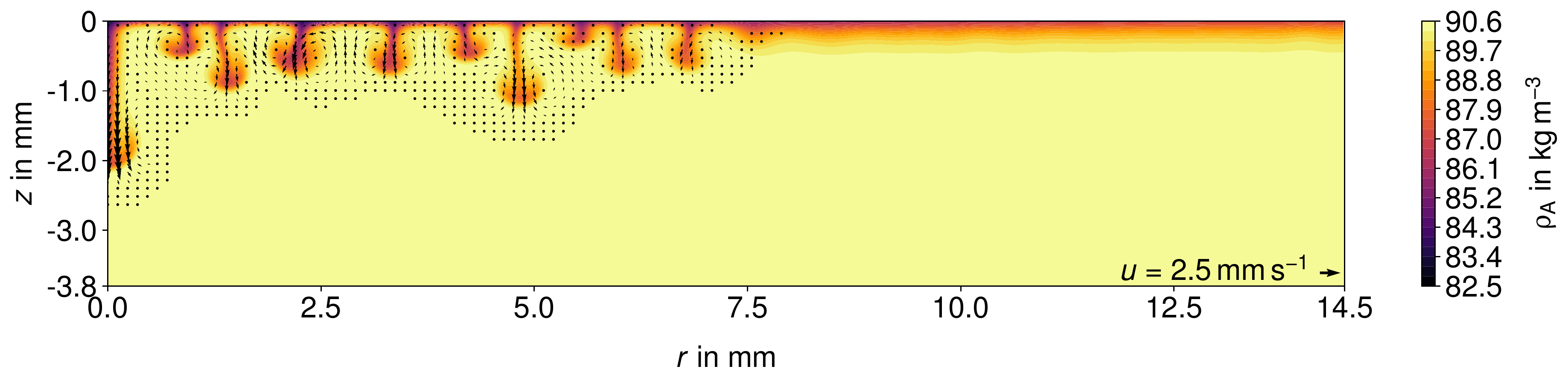}
        \end{subfigure}%
        \caption{\label{fig:c2D_t1} Instantaneous concentration and
                velocity fields in the positive electrode of a Li$||$Bi LMB for $I =
                \SI{1}{\ampere}$. (a) Uniform current distribution at $t =$
                \SI{7.5}{\second}
                ,
                (b) inhomogeneous current distribution at $t =$ \SI{6.5}{\second}.}
\end{figure*}

The flow shows a topology that is totally different from the organized motion typical for EVFs \cite{herreman2019numerical}. 
This is due to the absence of a ``basically stationary'' body force field ($\textbf{j}\times\textbf{B}$). 
The buoyancy term affecting the flow field shows an uneven and transient behavior
due to the strong coupling with the mass transport equation and its BCs.\\
We can qualitatively distinguish three different behaviors during compositional convection. 
Initially the concentration boundary layer starts to form
but the fluid remains at rest in a quasi stable regime \cite{foster1969onset},
it is diffusion dominated (Fig.~\ref{fig:dRhoA}, $\rho_{\text{A}}(t) \propto
\sqrt{t}$).
Then, in the second regime, the denser layer becomes unstable and plumes being
to detach from the top
interface; the penetration decelerates the heavy fluid,
generating the classical mushroom
shape, with vortex rings and more complex structures, as shown in
Fig.~\ref{fig:c2D_t1}.

The presence of an inhomogeneous current distribution affects time and
location of
the onset of the compositional convection. In this condition
instability starts in the center of the cell, as shown in
Fig.~\ref{fig:c2D_t1}b; the threshold limit for convection is
reached earlier (e.g. $\sim$\SI{2}{\second} for $I= $~\SI{1}{\ampere}) and the
peripheral
region remains fluid-dynamically stable and purely diffusive in the first
seconds. In the uniform case compositionally
driven motion begins everywhere affecting rapidly the top region of the layer,
see Fig.~\ref{fig:c2D_t1}a.
\begin{figure*}[t!]
        \centering
        \begin{subfigure}[]
                \centering
                \includegraphics[width=1.1\linewidth]{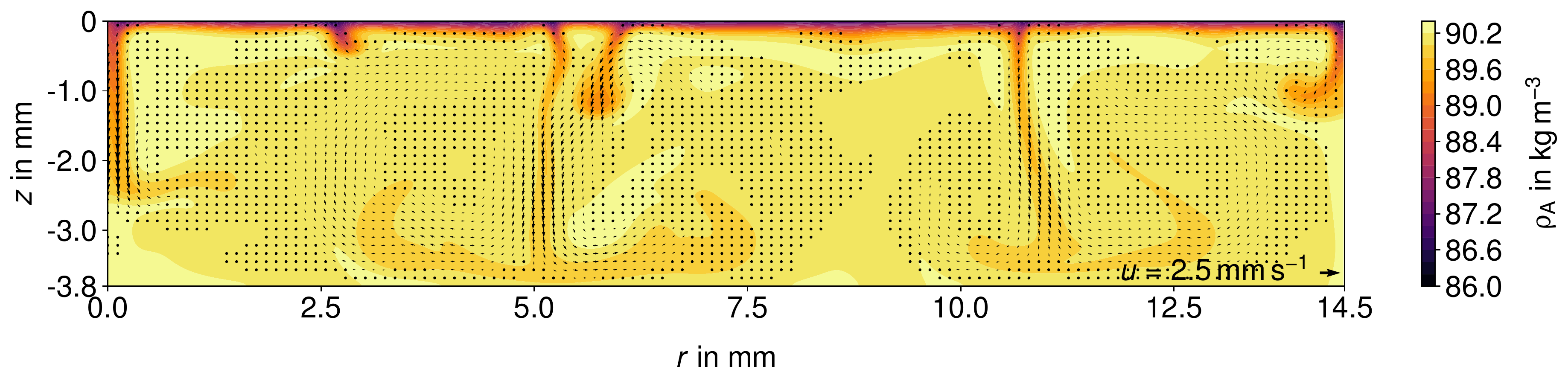}
        \end{subfigure}%
        \begin{subfigure}[]
        \centering
        \includegraphics[width=1.1\linewidth]{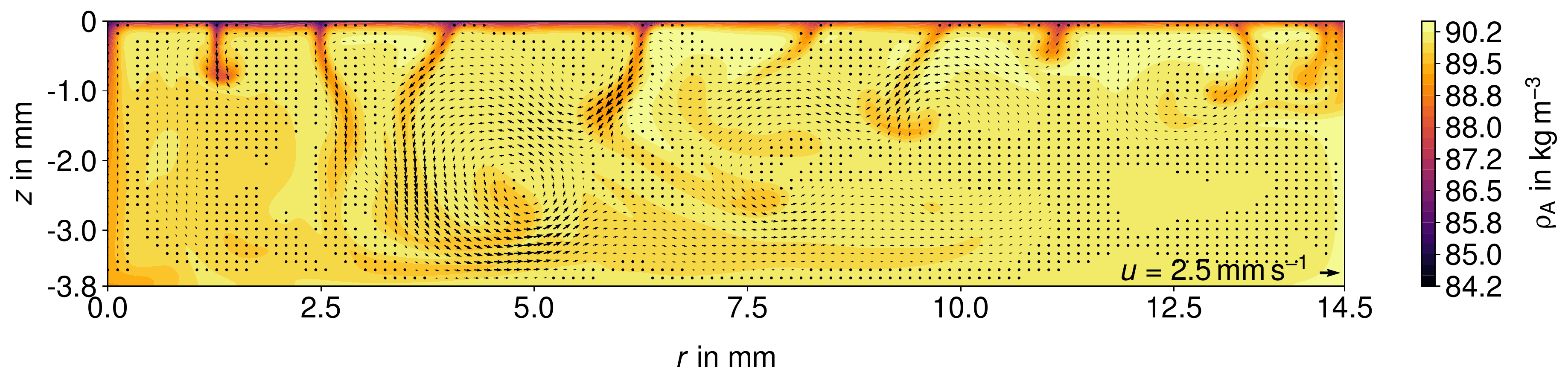}
    \end{subfigure}%
        \caption{\label{fig:c2D_t2} Instantaneous concentration and
                velocity fields in the positive electrode of a Li$||$Bi LMB for $I =
                \SI{1}{\ampere}$. (a) Uniform and (b) inhomogeneous current
                distribution at
                $t =$
                \SI{15}{\second}.}
\end{figure*}
When the
concentration perturbation has reached the bottom of the electrode, large scale
structures appear, as shown in Fig.~\ref{fig:c2D_t2};
the presence of walls becomes relevant and the acceleration of the flow
decreases. In this third regime the effect of the inhomogeneous current
distribution is less evident due to the presence of large scale unsteady and
chaotic structures that affect the full layer. If we analyse the evolution of
the difference between the average mass concentration at the interface and in
the volume $\Delta \rho_{\text{A}}$ we cannot observe relevant differences
between the uniform and inhomogeneous current distribution, especially in the
last
region when it becomes stable (Fig.~\ref{fig:dRhoA}). This concentration
difference is directly connected with the mass transport overpotential, as
described in \cite{personnettaz2019mass}, we can therefore conclude that the
inhomogeneous current distribution that we
studied has no relevant impact on the performance of the cell.

\section{Conclusions}
We have developed a numerical model able to describe mass
transport in the positive electrode of liquid
metal batteries, under different current distributions. In
the range of parameters studied we could numerically confirm that the different
electrodes areas, typical for experimental cells, have a negligible
impact on the cell performance during charge. They only affect fluid flow and
the concentration distribution in the first seconds when compositional
convection sets in.
A detailed investigation on a large range of operating parameters is currently
ongoing.
Further studies will be oriented to study the validity of the 2D
approximation with fully three-dimensional simulations and to investigate the
effect of the mechanical interaction with the electrolyte layer. 
Furthermore, the presence of a non-uniform current distribution has magneto-hydrodynamic implications that will be included 
in the future studies.

\Thanks{This work was supported by the Deutsche Forschungsgemeinschaft (DFG,
        German Research Foundation) by award number 338560565 and in frame of of
        the Helmholtz - RSF Joint Research Group ``Magnetohydro-
        dynamic instabilities: Crucial relevance for large scale liquid metal
        batteries and
        the sun-climate connection", contract No HRSF-0044 and RSF-18-41-06201.
        The computations
        were performed on the HPC-Cluster at the Center for Information Services
        and High Performance Computing (ZIH) at TU Dresden and on the HPC-Cluster
        ``Hemera" at Helmholtz--Zentrum Dresden -- Rossendorf.}


\bibliographystyle{mhd}
\bibliography{../../../references/references_ayt.bib}


\lastpageno	


\end{document}